\documentclass[aps,twocolumn,prl,amsmath,amssymb,floatfix,showpacs]{revtex4}
\usepackage{amsmath,amssymb,amsfonts,mathrsfs}
\usepackage{color}
\usepackage{epsfig}
\usepackage{psfrag}
\usepackage{graphicx}
\usepackage{color}

\begin{document}

\title{Topological Insulating Phases of Non-Abelian Anyonic Chains}

\author{Wade DeGottardi}
\affiliation{\small{Materials Science Division, Argonne National Laboratory, Argonne,
Illinois 60439, USA
}}

\date{\today}

\begin{abstract}
Boundary conformal field theory is brought to bear on the study of topological insulating phases of non-abelian anyonic chains. These topologically non-trivial phases display protected anyonic end modes. We consider antiferromagnetically coupled spin-1/2 su(2)$_k$ chains at any level $k$, focusing on the most prominent examples; the case $k = 2$ describes Ising anyons (equivalent to Majorana fermions) and $k = 3$ corresponds to Fibonacci anyons. We prove that the braiding of these emergent anyons exhibits the same braiding behavior as the physical quasiparticles. These results suggest a `solid-state' topological quantum computation scheme in which the emergent anyons are braided by simply tuning couplings of non-Abelian quasiparticles in a fixed network.
\end{abstract}

\maketitle

\emph{Introduction.}--- The notion of topological order has emerged as a powerful paradigm for the classification and discovery of new phases of matter~\cite{TQCA,TQCB,hasankane}. One manifestation of topological order is the existence of quasiparticles known as anyons. The exchange of two (abelian) anyons gives $\psi(r_2,r_1) = e^{i \theta} \psi(r_1,r_2)$ where $\psi$ is the many-body wave function describing the system and $\theta$ can (in principle) take any value; this is in stark contrast to the case of bosons ($\theta = 0$) and fermions ($\theta = \pi$)~\cite{TQCA,TQCB}. Even richer behavior arises in the case of non-Abelian anyons; the exchange of these objects enacts unitary transformations on the space of degenerate ground states~\cite{nayak}. This physics underlies \emph{topological quantum computation} (TQC) which proposes using a topologically ordered system as a robust quantum memory. The braiding of non-Abelian anyons has been suggested as a means of implementing fault-tolerant quantum gates in these systems~\cite{TQCA,TQCB,nayak}.

Interacting anyons are thought to exhibit a wide spectrum of behavior~\cite{anyonicqlo,anyonsuper,goldenchainA,trebst,itinerantanyons,longspin,ladder}. In this Letter, we apply boundary conformal field theory (BCFT) to the study of topological insulating phases of anyonic chains~\cite{bcftA,bcftAA,bcftB,bcftC}. That BCFT is useful in this context is natural since at its heart BCFT is a manifestation of the bulk-boundary correspondence and holography~\cite{entangle.entropy}. Although the method developed here is quite general, our analysis focuses on antiferromagnetically-coupled (AFM) spin-1/2 su(2)$_k$ anyonic chains at any level $k$, objects of study in non-Abelian Chern-Simons theories~\cite{TQCA,goldenchainA,trebst}. These models include two of the most prominent examples of non-Abelian anyons: Ising anyons which describe Majorana fermions ($k = 2$) and Fibonacci anyons for $k = 3$. The Fibonacci chain (also dubbed the `Golden chain') has attracted a great deal of interest with studies focusing on its behavior at criticality as well as the effects of disorder\cite{goldenchainA,goldenchainB,goldenchainC,topsymmetryA,topsymmetryB,disorderedgolden,trebst}. Fibonacci anyons capture the non-Abelian character of the quasiparticles in the $\mathbb{Z}_3$-parafermion ``Read-Rezayi'' state, a candidate theory for the $\nu = \frac{12}{5}$ fractional quantum Hall plateau~\cite{nayak}. These objects are of particular interest since, unlike Ising anyons, they are capable of performing \emph{universal} TQC~\cite{TQCA}.

\begin{center}
\begin{figure}
\includegraphics[width=8.0cm]{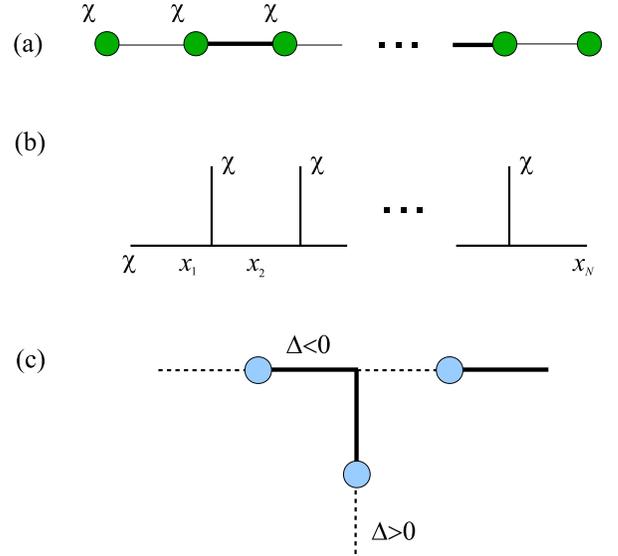}
\caption{(Color online) The degeneracy associated with (a) a chain of antiferromagnetically coupled su(2)$_k$  anyons (generically denoted by $\chi$) is lifted by their mutual couplings. For couplings $J_n = J\left(1 - (-1)^{n} \Delta \right)$ with $\Delta < 0$, the system is in a topological phase characterized by $\chi$ ($j = \frac{1}{2}$) boundary modes. The Hamiltonian of the chain (cf. Eq.~\ref{eq:ham}) is conveniently expressed in terms of (b) the fusion tree basis, where $x_n = 2j+1$ reflects the cumulative fusion product of the first $n$ anyons in the chain. (c) These emergent anyons may be manipulated by tuning the $\Delta$'s in a T-junction setup (see~\cite{tjunction}).}
\label{fig:setup}
\end{figure}
\end{center}

Our primary result is that an open chain of AFM spin-1/2 su(2)$_k$ anyons can exhibit a topologically non-trivial phase when the system is partially dimerized (see Fig. 1a). This behavior is well known in the context of Majorana fermions from the perspective of topological band theory and is also seen in a spin-$\frac{1}{2}$ Hiesenberg chain (related to the $k \rightarrow \infty$ limit)~\cite{kitaev,kitaevladder,endmodes}. Our theoretical method applies to all finite values of $k$. Starting with a description of the chain at criticality, we apply renormalization group (RG) arguments to track boundary degrees of freedom when a bulk gap is opened. In particular, the boundary degrees of freedom are shown to be an indicator of the topological properties of the chain. These theoretical predictions are verified with numerical diagonalizations of the Ising and Fibonacci chains.

This phenomenon suggests a `solid-state' version of TQC in which emergent anyons can can be manipulated in a T-junction network of coupled quasiparticles (Fig. 1c)~\cite{tjunction}. Building on work in~\cite{tjunction}, we sketch a general proof that the emergent anyons exhibit the same braiding properties as the physical quasiparticles. This scheme has the virtue of not requiring the large-scale motion of the quasiparticles composing the network~\cite{TQCA,TQCB,nayak,simon,footnote1}.

\emph{Interacting anyonic chains.}--- We consider a linear array of $N$ non-Abelian anyons (generically denoted $\chi$) (see Fig. 1b) with $N$ even. For $N \gg 1$, these anyons encode a $\sim d_\chi^N$-dimensional subspace, where $d_\chi$ is the \emph{quantum dimension} of $\chi$~\cite{nayak}. For adjacent anyons separated by a distance $D$ with $D \lesssim \xi$ where $\xi$ is the correlation length of the system, this degeneracy is lifted~\cite{TQCA}. We consider AFM-coupled spin-1/2 anyons belonging to su(2)$_k$ for finite $k$~\cite{TQCA,TQCB}. For a given $k$, these theories contain anyon types labeled by $j = 0, \frac{1}{2}, 1,...,\frac{k}{2}$. We consider chains formed by $j = \frac{1}{2}$ anyons. The fusion rules of this theory obey an analog of angular momentum addition, i.e. $j_1 \times j_2 = |j_1 - j_2| + (|j_1 - j_2| + 1) + ... + \min ( j_1 + j_2 , k - j_1 - j_2)$, thus $j = 0, \frac{k}{2}$ correspond to the identity $I$. Akin to a Heisenberg spin interaction, the interaction energy between anyons depends on their mutual fusion product~\cite{goldenchainA,goldenchainB,goldenchainC}. Here we consider a chain with nearest-neighbor couplings described by the Hamiltonian
\begin{equation}
H = \sum_{n} J_{n}^{\phantom{(}} \Pi_{n}^{(I)},
\label{eq:ham}
\end{equation}
with $J_n < 0$ and where $\Pi_{n}^{(I)} = | I \rangle \langle I |$ is the projector onto the trivial (identity) fusion channel for the anyons connected by the $n^{\textrm{th}}$ link (see Fig. 1a)~\cite{goldenchainA,trebst,hash}.

In order to describe the collective state of a chain, we employ the fusion tree basis~\cite{goldenchainA,TQCA}. The transformations between the local and fusion tree basis can be accomplished by unitary transformations described by so-called $F$-matrices (see Table I). Each degree of freedom $x_n$ (see Fig. 1b) indicates the cumulative fusion of the first $n$ anyons (on the left-hand side of the chain). Letting $x_n = 2j+1$ where $j$ is the corresponding anyon, Eq.~\ref{eq:ham} is equivalent to the critical $p = k+2$ restricted solid-on-solid (RSOS) model ~\cite{goldenchainA,trebst,CFT}. In the RSOS model, each integer $x_n$ corresponds to a height with ($1 \leq x_n < p$) subject to the condition $| x_n - x_{n+1} | = 1$, which in this context enforces the fusion rules. Since $x_1$ represents a $j = \frac{1}{2}$ anyon, we take $x_1 = 2$. The quantity $x_N$ represents the total fusion product of the chain.

\begin{table}
\caption{Conformal data for Ising and Fibonacci theories. The $\left(s^a\right)$ are Pauli spin matrices $a = x, y, z$ which convert from the local fusion basis to the fusion tree basis.}
\begin{tabular}{|c||l|l|}
\hline
\hline
    & Ising & Fibonacci \\
\hline
  fields & $\sigma$, $\psi$, $I$ & $\tau$, $I$ \\
  $d_\chi$ & $d_\sigma = \sqrt{2}$  & $d_\tau = \frac{1+\sqrt{5}}{2}$  \\
  F-matrices & $F_\sigma^{\sigma \sigma \sigma} = \frac{1}{\sqrt{2}}\left( s^z + s^x \right)$  & $F_\tau^{\tau \tau \tau} = \frac{1}{\varphi} s^z + \frac{1}{\sqrt{\varphi}} s^x$ \\
  fusion rules & $\sigma \times \sigma = I + \psi$ & $\tau \times \tau = I + \tau$ \\
  & $\sigma \times \psi = \sigma$ & \\
  RSOS map & $I \rightarrow 1$, $\sigma \rightarrow 2$  & $I \rightarrow 4/1$, $\tau \rightarrow 2/3$ \\
  & $\psi \rightarrow 3$ & for $n$ odd/even \\
  \hline
\end{tabular}
\end{table}

We now give the Hamiltonians for $k = 2,3$. The Ising model fusion rules (Table I) dictate that (see Fig. 1c) $x_n = 2 (\sigma)$ for $n$ odd and $x_{n} = 1 (I),3 (\psi)$ for $n$ even~\cite{CFT}. We shift Eq.~\ref{eq:ham} by an overall constant giving $H = \frac{1}{2} \sum_{n = 1}^N  J_n \left( \Pi^{(I)}_n - \Pi^{(\psi)}_n \right)$. Applying the $F$-matrix in Table I, we obtain
\begin{equation}
H = \frac{1}{2} \sum_n J_{2n} s_{n}^x + J_{2n+1} s_{n}^z s_{n+1}^z,
\label{eq:Hising}
\end{equation}
the quantum Ising model~\cite{kitaev}. Equation~\ref{eq:Hising} is described in terms of pseudospins $I = | \downarrow \ \rangle_z$, $\psi  = | \uparrow \ \rangle_z$ (for $N$ Ising anyons, there are $N/2$ spins). This system can be mapped to a spinless $p$-wave superconductor via a Jordan-Wigner transformation, i.e. $\Psi_n = \prod_{j} \sigma_j^x \left( \sigma^x_j + i \sigma^y_j \right)$~\cite{kitaev,kitaevladder}.

For Fibonacci anyons, the state of the system is characterized by  $x_n = 1 (I), 2 (\tau)$ with $x_1 = 2 (\tau)$. The Hamiltonian (in the fusion tree basis) is given by $H = J_1 f_1 + \sum_{n = 2}^{L-1} J_n H_n$ where
\begin{eqnarray}
H_n &=&  \left(1 - f_{n-1} - f_{n+1} \right) \\ && + f_{n-1}f_{n+1} \left( \varphi^{-3/2} s_n^x + \varphi^{-3} f_n + 1 + \varphi^{-2} \right), \nonumber
\label{eq:hamfib}
\end{eqnarray}
and $f_n = \frac{1}{2} ( 1 - s^z_n)$ is the $\tau$-occupation number on the $n^{\textrm{th}}$ link~\cite{goldenchainA} and we have used pseudospins $I = | \uparrow \ \rangle_z$, $\tau = | \downarrow \ \rangle_z$. Note that two adjacent $I$'s in the fusion tree violate the fusion rules.

A remarkable aspect of these chains is that for uniform couplings $J_n = J$ the system is critical~\cite{goldenchainA,goldenchainB,goldenchainC,topsymmetryA,topsymmetryB}. The critical RSOS model is described by the minimal CFT model $\mathcal{M}(k+2,k+1)$, with central charge $c = \frac{(k+4)(k-1)}{(k+1)(k+2)}$~\cite{CFT}. This model possesses primary fields $\phi_{(r,s)}$ for $0 \leq r \leq k+1$ and $0 \leq s \leq k+2$, with the identification $\phi_{(r,s)} \equiv \phi_{(k-r+1,k-s+2)}$. The critical quantum Ising model is described by $\mathcal{M}(4,3)$ with primary fields $\phi_{(1,1)} \ (I, 0)$,  $\phi_{(1,2)} \ (\sigma,\frac{1}{16})$, $\phi_{(2,1)} \ (\psi, \frac{1}{2})$. Here, the scaling fields $\phi_{(r,s)}$ are in 1-to-1 correspondence with the Ising anyons in Table I. However, there are 6 scaling fields for the tricritical Ising model  $\phi_{(1,1)} \ (I, 0)$, $\phi_{(2,1)} \ (\sigma',\frac{7}{16})$, $\phi_{(3,1)}/\phi_{(1,4)} \ (\varepsilon'', \frac{3}{2})$, $\phi_{(1,2)} \ (\varepsilon, \ \frac{1}{10})$, $\phi_{(1,3)} \ (\varepsilon', \frac{3}{5})$, $\phi_{(2,3)} \ (\sigma, \frac{3}{80})$~\cite{CFT,goldenchainA}. The connection between these fields and $\{I,\tau\}$ can be made by making use of the topological symmetry~\cite{goldenchainA,goldenchainB,topsymmetryA}. For general $k$, the parameter $s$ controls the topological sector to which $\phi_{(r,s)}$ belongs: the field is topologically trivial if $s = 1$ and thus corresponds to $I$. Topologically non-trivial fields are given by $r = 1$ and $s \neq 1, k+1$ and correspond to $j$, with $j \neq \frac{1}{2}, \frac{k}{2}$~\cite{goldenchainC}. For $k=3$, this gives $\{I, \sigma', \varepsilon''\} \rightarrow I$ while $\{ \varepsilon, \varepsilon', \sigma \} \rightarrow \tau$.

\begin{center}
\begin{figure}
\includegraphics[bb = 34 383 579 600,width=9.0cm,height=4.0cm]{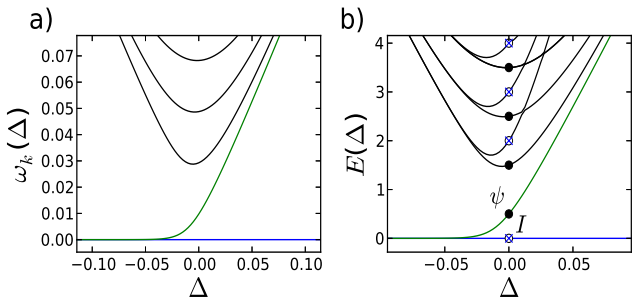}
\caption{(Color online) (a) Single-particle spectrum and (b) the many-body spectrum of the coupled Ising anyons described by Eq.~\ref{eq:Hising} for ($N = 100$). In both graphs, the energy has been rescaled so the first excited state matches scaling dimension $\frac{1}{2}$ of $\psi$. The system contains end modes $\sigma \times \sigma = I + \psi$. The state corresponding to $I$ (at $E=0$) and those corresponding to its descendants (at $E = 2,3,...$) are marked by a circle with a cross. The states corresponding to $\psi$ and its descendants (at $E = \frac{1}{2} + 1, \frac{1}{2} + 2,...$) are indicated by a filled dot.}
\label{fig:ising}
\end{figure}
\end{center}

\emph{Topological Insulating Phases}--- We now consider insulating phases of these systems. The scaling fields of $\mathcal{M}(k+2,k+1)$ have momentum $K = 0$ or $K = \pi$~\cite{goldenchainC}; textures of the form $J_n = J \left( 1 - (-1)^{n} \Delta \right)$ open a gap in the spectrum~\cite{goldenchainB,goldenchainA} ($J<0$). In the extreme limit $\Delta = -1$ (see Fig. 1a), the chain is composed of dimers (which fuse to $I$) and two isolated anyons at each end of the chain. We expect an end state will persist as long as the bulk gap does not close~\cite{kitaev}. Thus, we anticipate that $\Delta < 0$ represents a topologically non-trivial phase while for $\Delta > 0$ the system is trivial. While these arguments are suggestive, we seek a more systematic theoretical approach that can identify the topological phases near criticality ($|\Delta| \ll 1$).

We have performed numerical diagonalizations of both the Ising and Fibonacci chains (see Figs. 2 and 3). Figure 2a shows the eigenenergies of the $p$-wave superconductor corresponding to Eq.~\ref{eq:Hising}, while Fig. 2b shows the corresponding many-body spectrum. The ferromagnetic phase ($\Delta < 0$) corresponds to a topologically non-trivial phase in the superconductor~\cite{kitaev,kitaevladder}. The topological phase hosts end states and the nearly degenerate ground state is split by the residual interaction $J_{\textrm{eff}} \sim J e^{- \Delta L}$ between these end modes~\cite{kitaev,kitaevladder}. A key feature of the results in Fig. 3a for the Fibonacci chain is that there exists a similar degeneracy as well, suggesting the existence of two end modes in that system as well.

We now apply BCFT and RG arguments to show that as suggested above, these results hold for general $k$, i.e. that such chains will exhibit $j = \frac{1}{2}$ end modes for $\Delta < 0$. For $\Delta = 0$, the chain is critical and for an open chain is described by the action $\mathcal{S}_0 = \mathcal{S}_{0,\textrm{bulk}} + \mathcal{S}_{0,R} + \mathcal{S}_{0,L}$. The term $\mathcal{S}_{0,\textrm{bulk}}$ represents the bulk action of the minimal model $\mathcal{M}(k+2,k+1)$ while $\mathcal{S}_{0,L/R}$ describes the left($L$)/right($R$) ends of the fusion tree. The spectrum of a critical finite size chain is (up to an overall shift and rescaling) given by the conformal dimensions of the fields appearing in the fusion product of $\mathcal{S}_{0,R}$ and $\mathcal{S}_{0,L}$ along with the corresponding descendants~\cite{goldenchainA,CFT}. It should be emphasized that this is a continuum description of the fusion tree basis; $\mathcal{S}_{0,R}$ represents the total fusion product of the system and \emph{not} a physical end mode.

For the Ising case, we have $\mathcal{S}_{0,L/R} = \int dt \; \sigma_{L/R} (t)$ and the spectrum is described by $\sigma \times \sigma = I + \psi$ plus descendants (see Fig. 2b). In the case of the RSOS model $p>4$ ($k > 2$), the boundary field $\phi_{(1,s)}$ arises when an end is fixed to $x_N = s$ but the penultimate degree of freedom $x_{N-1}$ is unconstrained; the field $\phi_{(r,1)}$ arises from fixing an end to $x_N = r$ and $x_{N-1} = r+1$ (or vice versa)~\cite{RSOSbcsA,RSOSbcsB,tricritbcsA,tricritbcsB}. While in most physical settings we expect that the total fusion product is fixed, in determining the end mode structure it is convenient to temporarily relax this condition. A long chain's total fusion can take on values $x_N = 1, 3, 5,... \leq k+2$. Since at criticality there is no constraint on the penultimate site $x_{N-1}$, for $x_N = s$ we have
\begin{equation}
\mathcal{S}_{0,R} = \lambda_s \int dt \, \phi_{(1,s)}(t).
\label{eq:rboundary}
\end{equation}
For a chain with an unconstrained total fusion, the spectrum corresponds to the tensor product of all possible values of $s$ along with the corresponding descendent fields. For example, in the Fibonacci case we have $\mathcal{S}_{0,R} = \lambda \int dt \ I(t) / \varepsilon'(t)$ depending on whether the chain fuses to $I$ or $\tau$, respectively; $\varepsilon'$ is a topologically non-trivial field and therefore a proxy for $\tau$. For $k > 2$, the constraint $J_1$ is not included in the mapping to the RSOS model and must therefore be treated as a boundary condition. This gives $(x_1,x_2) = (2,1)$ with $\mathcal{S}_{0,L} = \lambda \int dt \; I(t)$. These identifications are consistent with the critical spectra in Fig. 3a and this accounts for CFT field assignments in~\cite{goldenchainA}. A Fibonacci chain with a constrained fusion product $x_N = 1 (I)$ is shown in Fig. 3b.

We now consider the effect of non-zero $\Delta$ ($|\Delta| \ll 1$) which can induce both bulk and boundary perturbations. Only topologically trivial fields arise as perturbations in the bulk~\cite{goldenchainA,topsymmetryA,topsymmetryB}. The field $\Phi_{(2,1)}(z,\bar{z}) = \phi_{(2,1)}(z) \otimes \phi_{(2,1)}(\bar{z})$ is the only topologically trivial RG relevant field (its scaling dimension is $h = \frac{k+4}{2(k+1)} < 2$~\cite{goldenchainC}).

For a gapped system, the possible low-energy fixed points $\mathcal{S}_{R}^\ast$ give the fusion products of the end modes. From this information the physical end modes of the system can be identified. While in general it is possible for bulk fields to induce boundary RG flows~\cite{bulkboundaryRG}, this is not the case for the class of anyons studied here~\cite{bulkboundarynote}. The fields which can appear in $\mathcal{S}_{0,R}$ (Eq.~\ref{eq:rboundary}) correspond to boundary conditions with $x_N = r$ and freedom in the penultimate site $x_{N-1} = r \pm 1$. The boundary perturbation associated with any $\Delta < 0$ does not lift this degeneracy. Thus, for $k > 2$ we expect that from the BCFT rules of the RSOS model, the low energy fixed points $\mathcal{S}_{L}^\ast$ will consist of scaling fields $\phi_{(1,s)}$ which appear in $\mathcal{S}_{0,R}$ \emph{and} are RG relevant (with conformal dimension $ < 1$). For $k = 2$, this gives $\mathcal{S}_{R/L}^\ast = \lambda \int dt \ \sigma(t)$. For $k > 2$, this gives $\mathcal{S}_L^\ast \propto \int dt \ I$ and $\mathcal{S}_{R}^\ast = \lambda \int dt \ I / \phi_{(1,3)}(t)$ (the fields $\phi_{(1,s)}$ with $s > 3$ are irrelevant for all $k$). These expectations are confirmed by the results of boundary RG flows for the Ising and tricritical Ising models~\cite{bcftA,bcftAA,bcftB,bcftC,exactRG}, though we expect these arguments hold for all $k$. In all cases, for $\Delta < 0$ the fusion products of the fields in $\mathcal{S}_{R}^\ast$ and $\mathcal{S}_L^\ast$ are consistent with two $j = \frac{1}{2}$ end modes which can fuse to $\frac{1}{2} \times \frac{1}{2} = 0 + 1$. We thus infer that the system is topologically non-trivial for $\Delta < 0$.

For $\Delta > 0$, it is clear from Eqs.~\ref{eq:Hising} and \ref{eq:hamfib} (and holds for all $k$) that the degeneracy associated with $x_{N-1} = r \pm 1$ is lifted. This physics is nicely illustrated by the Ising case. It it is clear from Eq.~\ref{eq:Hising} that when $\Delta > 0$ the term $s_{n}^x$ is dominant over $s_{n}^z s_{n+1}^z$ and this is true at the ends of the system in particular. Thus, the boundary perturbation which arises from taking $\Delta > 0$ is tantamount to the introduction of a `magnetic field' at the ends of the system. This leads to the stable fixed point $\mathcal{S}^\ast_{R} = \frac{1}{\sqrt{2}} \int dt \left[ I(t) + \psi(t) \right]$ corresponding to boundary spins pointing in the $+\hat{x}$-direction. We expect for all $k$ that the perturbation associated with $\Delta > 0$ will fix $x_{N-1}$ and thus, from the BCFT rules for the RSOS model, will induce RG flow to a fixed point described by a superposition of topologically trivial fields, i.e.
\begin{equation}
\mathcal{S}_{R}^\ast =  \sum_{r} c_r \int dt \ \phi_{(r,1)}.
\end{equation}
That $\mathcal{S}_{R}^\ast$ is equivalent to $I$ requires that the end modes of the system are trivial as well.  Again we expect this to hold for all $k$, however the boundary RG has been worked out only (as far as we are aware) for the Ising and tricritical Ising models (corresponding to $k = 2,3$) ~\cite{bcftA,bcftAA,bcftB,exactRG}.

\begin{center}
\begin{figure}
\includegraphics[bb = 34 383 579 600,width=9.0cm,height=4.0cm]{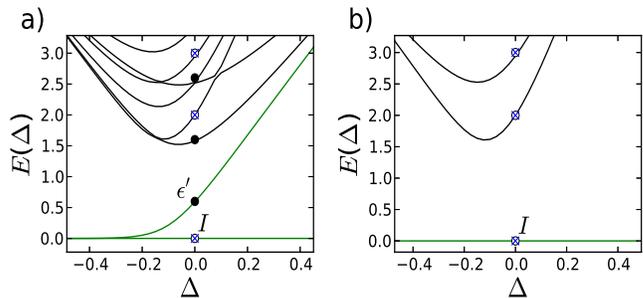}
\caption{(Color online) Low-energy spectrum of the AFM `golden chain' for (a) $N = 18$ Fibonacci anyons with an unconstrained fusions product. As in Fig. 2b, the energies have been rescaled to match the dimension of the corresponding primary fields. The states corresponding to $I$ ($E=0$) and its descendants ($E = 2,3,...$) are indicated by a circle with cross. The states corresponding to $\varepsilon'$ and its descendents at energies $E = \frac{3}{5}, \frac{3}{5}+1,...$ are marked by a filled dot. (b) is the spectrum of a system with a total fusion product constrained to be $I$ ($x_N = 1$).}
\label{fig:fibonacci}
\end{figure}
\end{center}

\emph{Topological Quantum computation with emergent anyons.}--- Consider a T-junction setup shown in Fig. 1c~\cite{tjunction}. Slowly varying textures of $\Delta$ will give rise to protected anyonic states localized near domain walls (where $\Delta$ passes through zero). The pressing question is whether the exchange of the emergent anyons will give rise to non-trivial unitary transformations in the degenerate ground state subspace they encode. This is an important issue since the results of anyonic braiding depend on the so-called $R$-matrix~\cite{TQCA} which in turn depends on the `spin' of the anyons and therefore ostensibly this operation references a physical rotation. However, it was shown in~\cite{tjunction} that exchanges in a T-junction setup \emph{do} enact the usual unitary transformations for Majorana fermions.

We argue that this result holds for all types of anyons. This follows from the tight constraints exhibited by anyonic theories -- a feature known as Oceanu rigidity~\cite{TQCA,TQCB,nayak}. In a topological phase, the possible fixed points $\mathcal{S}_{R}^\ast$ can be used to determine the identity of the end modes. The identity of the end modes determines the relevant $F$-matrices since these depend on the Hilbert space formed by the anyons and not on the geometry of the system~\cite{nayak}. The unitary transformation (braiding) enacted by the physical rotation of two anyons depends on both the $F$- and $R$-matrices. Now, the $F$- and $R$-matrices obey the hexagon identity~\cite{TQCA,TQCB} which uniquely determines the $R$-matrix given the $F$-matrix~\cite{TQCA}. Crucially, the only feature of the $R$-matrix that the hexagon identity assumes is that it enacts the exchange of two anyons. Thus, we conclude that the exchange of two emergent anyons satisfies the hexagon identity and thus leads to the usual $R$-matrix and braiding properties.

\emph{Conclusions}--- We have studied the topological insulating phases of anyonic chains. BCFT has been shown to be a powerful tool in the study these systems. The physics studied here suggests a TQC scheme in which emergent anyons are braided and manipulated, requiring only the fine tuning of the couplings in an otherwise fixed array of physical quasiparticles. Other types of anyons hold great promise for even richer behavior. Ferromagnetically-coupled Fibonacci anyons are described by the $\mathbb{Z}_3$ parafermion theory which has fields of momenta $K = 0, \frac{2\pi}{3}, \frac{4\pi}{3}$~\cite{goldenchainC}. Thus, the relatively simple dimerization picture which applies for the AFM case will no longer hold. Higher spin versions of these chains represent another direction for future study.

For support, I thank UChicago Argonne, LLC, operator of Argonne National Laboratory, under contract No. DE-AC02-06CH11357. I would like to thank Diptiman Sen, Smitha Vishveshwara, Konstantin Matveev, Brian Skinner, and Ivar Martin for their insightful comments and suggestions.

\appendix


\begin{thebibliography}{99}
\bibitem{footnote1} The dimerized golden chain was briefly discussed in Ref.~\cite{goldenchainB}).
\bibitem{bulkboundarynote}  The boundary fields which could be induced by $\phi_{(2,1)}$ are those appearing in the fusion product $\phi_{(2,1)} \times \phi_{(2,1)}$. Since all such fields are irrelevant, boundary flows can only occur from boundary perturbations.
\end{thebibliography}
\end{document}